\documentclass[showpacs,prl,twocolumn,superscriptaddress]{revtex4}
\usepackage{graphicx}
\newcommand{\etal}{{\it et al.}}

\begin{document}

\title{Dichroism as a probe for parity-breaking phases of spin-orbit coupled metals}

\author{M. R. Norman}
\affiliation{Materials Science Division, Argonne National Laboratory, Argonne, IL 60439, USA}

\begin{abstract}
Recently, a general formalism was presented for gyrotropic, ferroelectric, and
multipolar order in spin-orbit coupled metals induced by spin-spin interactions.  Here, I point out that
the resulting order parameters are equivalent to expectation values of operators that determine 
natural circular dichroic signals in optical and x-ray absorption.
Some general properties of these operator equivalents and the resulting dichroisms are mentioned,
and I list several material examples in this connection, including Weyl semimetals.
The particular case of the tensor order in
the pyrochlore superconductor Cd$_2$Re$_2$O$_7$ is treated in more detail,
including calculations of the x-ray absorption and circular dichroism at the O K edge.
\end{abstract}

\date{\today}

\maketitle

Reciprocal (natural) and non-reciprocal dichroism, and analogous effects in resonant x-ray scattering,
are well known probes of materials which break
inversion or time-reversal symmetry, respectively.  There has been a resurgence of interest
in these effects due to their prominent observation in multiferroics and chiral magnets.
It has been recently realized that similar manifestations occur in novel topological materials.
For instance, circular dichroism due to the chiral anomaly in Weyl semimetals has been
proposed by Hosur and Li \cite{hosur} and observed by Kerr rotation in Cd$_3$As$_2$ \cite{zhang}.
This is effected by applying parallel electric and magnetic
fields in order to realize the chiral $E \cdot B$ term, which has been earlier studied in
the magnetoelectric material Cr$_2$O$_3$ \cite{hehl}.

Much of the interesting phenomena associated with topological materials can be traced
to momentum-spin locking due to spin-orbit coupling.
Fu has pointed out that spin-spin interactions in the presence of
spin-orbit coupling not only lead to magnetism, but other types of order due to this same
momentum-spin locking \cite{fu}.  This gives rise to the free-energy
\begin{eqnarray}
E_{ss} & = & \sum_{k,k'} F_0 s(k) \cdot s(k') + F_1 (k \cdot s(k)) (k' \cdot s(k')) \\
& & + F_2 (k \times s(k)) (k' \times s(k')) + F_3 Q_{ij}(k) Q_{ij}(k') \nonumber
\end{eqnarray}
where the quadrupolar term is
\begin{equation}
Q_{ij}(k) = \frac{1}{2} (k_i s_j(k) + k_j s_i(k)) - \frac{1}{3} (k \cdot s(k)) \delta_{ij}
\end{equation}
with $k$ the momentum and $s$ the spin.
For each term, one can specify a particular order - ferromagnetism for $F_0$, isotropic
gyrotropic order for $F_1$, ferroelectricity for $F_2$, and pseudo-deviator order for $F_3$ - the last has been suggested
for the pyrochlore superconductor Cd$_2$Re$_2$O$_7$ \cite{petersen}.  Besides $F_0$, these all break
inversion symmetry, noting that a pseudo-deviator is a rank-2 tensor that is odd under inversion.
Moreover, as also discussed in Ref.~\onlinecite{fu}, each term leads to a unique splitting of the Fermi surface 
(for some illustrations of this splitting, see Ref.~\onlinecite{fujimoto}).

Here, I wish to point out that the resulting operators listed above ($k \cdot s(k)$, etc.) are equivalent to operators that have
been suggested to determine optical and x-ray dichroism \cite{goulon,marri} if a mapping is made between $k$ and $t$,
where $t$ is the polar toroidal moment \cite{tugushev,spaldin}.  $k$ and $t$ have the same transformation properties
under inversion and time reversal (both being odd) \cite{schmid}, and thus for the argumentation made in Refs.~\onlinecite{goulon}
and \onlinecite{marri}, they are equivalent.
This can also be appreciated from the analogy between the spin-orbit interaction, $(k \times \nabla V) \cdot s$, and the
toroidal contribution to the free energy, $-t \cdot (P \times M)$, where $V$ is the crystal potential, $P$ the polarization,
and $M$ the magnetization.
Polar toroidal ordering has been seen in a number of multiferroics \cite{fiebig,kubota}, and is also the
basis for a novel theory of the pseudogap phase of cuprates by Varma and collaborators \cite{varma}, following earlier theoretical 
suggestions of Gorbatsevich \etal~\cite{gorb}.
In general, there are several operator equivalents that can describe natural dichroism as listed in Table 1, including a recently
advocated form involving the Berry curvature by Zhong, Orenstein and Moore \cite{zhong}.

\begin{table}
\caption{Various operator equivalents for natural circular dichroism.  These are formed from the vectors listed in the first two columns,
with the transformation properties of these vectors under inversion and time-reversal (respectively) listed in parenthesis.
The scalar product of the two is responsible for pseudo-scalar optical activity,
and can be shown to vanish in the single particle approximation because of the orthogonality of the two vectors.
The cross product of the two is responsible for polar vector optical activity.
The remaining five components involving the two vectors is a traceless symmetric matrix, responsible for pseudo-deviator activity
that dominates in the x-ray regime.
Here, $m$ is the magnetization, $t$ the polar toroidal moment, $d$ the electric dipole moment, $g$ the axial toroidal moment, $v$
the velocity, $\Omega_B$ the Berry curvature, and $k$ the momentum.
}
\begin{ruledtabular}
\begin{tabular}{ccc}
Vector 1 & Vector 2 & Reference \\
\colrule
m (+,-) & t (-,-) & \cite{goulon,lovesey2} \\
d (-,+) & g (+,+) & \cite{sergio,sergio3} \\
v (-,-) & $\Omega_B$ (+,-) & \cite{zhong} \\
k (-,-) & m (+,-) & this work \\
\end{tabular}
\end{ruledtabular}
\end{table}

\begin{table}
\caption{Examples of several non-centrosymmetric materials and the types of natural optical activity that can occur \cite{jc}, where PS stands
for pseudo-scalar, PV for polar vector, and PD for pseudo-deviator (with 1 denoting allowed).
Just because it is allowed does not necessarily mean dichroism can occur.  For instance, for Cd$_2$Re$_2$O$_7$ (I$\bar{4}$m2
space group), TaAs, and Bi2212, XNCD is zero for x-rays propagating along the c axis, due to the presence of glide and mirror planes.
}
\begin{ruledtabular}
\begin{tabular}{ccccc}
 & Space Group & PS & PV & PD \\
\colrule
HgTe & F$\bar{4}$3m & 0 & 0 & 0 \\
BiTeI & P3m1 & 0 & 1 & 0 \\
TaAs & I4$_1$md & 0 & 1 & 0 \\
SrSi$_2$ & P4$_3$32 & 1 & 0 & 0 \\
Cd$_2$Re$_2$O$_7$ & I4$_1$22 & 1 & 0 & 1 \\
 & I$\bar{4}$m2 & 0 & 0 & 1 \\
 LiOsO$_3$ & R3c & 0 & 1 & 0 \\
 CdS, ZnO & P6$_3$mc & 0 & 1 & 0 \\
 Te & P3$_1$21 & 1 & 0 & 1 \\
 Bi$_2$Sr$_2$CaCu$_2$O$_{8+x}$ & Bb2b & 0 & 1 & 1 \\
\end{tabular}
\end{ruledtabular}
\end{table}
 
To understand the consequences of this, I discuss each order parameter in turn, denoting each as $X_i$, where $i$ is the index of
$F_i$ in Eq.~1 (that is, $X_0 \equiv \sum_k s(k)$, $X_1 \equiv \sum_k k \cdot s(k)$, etc.).  $X_0$ is ferromagnetism,
and can be determined by XMCD (x-ray magnetic circular dichroism) \cite{xmcd}.
The rest, though, are associated with natural (as opposed to non-reciprocal) optical activity, and correspond to HHE (piezo-electric) terms in the free
energy, where H and E are the magnetic and electric fields \cite{goulon}.
$X_1$ is a pseudo-scalar
(also known as an axial toroidal monopole \cite{sergio3}).  It can be seen
in natural optical dichroism, due to an interference between electric dipole (E1) and magnetic dipole (M1) terms, and the condition for
its observation for various point groups has been tabulated by Jerphagnon and Chemla \cite{jc}, with a few relevant examples shown in Table 2.
Recently, natural dichroism has also been seen in connection with spin wave excitations in a field induced chiral magnet, where it is due to dynamic terms in the
diagonal part of the magnetoelectric susceptibility tensor \cite{bordacs}.  In general, one expects
natural dichroism to be present for chiral magnets since $S \times S \equiv i S$, leading to a $k \cdot S$ term in scattering \cite{lovesey}.
In the x-ray regime,
it is difficult to observe since the M1 matrix elements for excitations out of core orbitals are typically very small (though they can be significant
for shallow core levels).

Moreover, Marri \etal~\cite{marri2} have pointed out that the pseudo-scalar vanishes in the single-particle
approximation.  This can be appreciated from Table 1, where $m$ and $t$ are orthogonal vectors ($t$ being the cross product of $r$ and $m$), meaning
that they must refer to different electrons to obtain a non-zero result.  The same logic \cite{sergio} also applies to $d$ and $g$ ($g$ being the 
cross product of $r$ and $d$).  The same result has recently been pointed out in the Berry curvature approach by Zhong \etal~\cite{zhong}, since again,
$v$ and $\Omega_B$ are orthogonal vectors ($v$ being the cross product of $E$ and $\Omega_B$).
The author has verified this using the FDMNES program (described below) for tellurium (Table 2),
where the E1-M1 (electric dipole - magnetic dipole) interference terms are found to be traceless for the various edges studied (L$_1$, L$_{2,3}$, M$_{4,5}$, N$_{4,5}$),
noting that the E1-E2 (electric dipole - electric quadrupole) interference terms are traceless by definition.
Obviously, the single-particle approximation cannot be valid in general, since natural optical activity can be
seen in liquids where the pseudo-deviator averages out \cite{barron}.

As commented in Ref.~\onlinecite{fu}, $X_2$ is equivalent to ferroelectric order.  The dichroism
associated with this is the vector part of the natural optical activity, commonly known as Voigt-Fedorov dichroism.
It is due to a longitudinal component of the electric polarization, a specific property of pyroelectric materials \cite{jc}.
Typically, the small longitudinal electric field is difficult to observe, but one can easily study a related effect where the polarization of the reflected beam rotates
if the propagation vector of the incoming beam is not along the surface normal \cite{graham}.
This effect has been seen optically in CdS (Table 2) \cite{cds}.  Recently, it has also been
seen by resonant x-ray scattering in ZnO as an x-ray circular intensity differential \cite{zno}.
Since the latter measurements were done at the Zn K edge, the resulting operator
actually involves the orbital and orbital toroidal moments rather than their spin counterparts \cite{goulon}.
A related non-reciprocal effect for generating a longitudinal electric field in topological Weyl semimetals that violate time-reversal symmetry
has been pointed out by Kargarian \etal~\cite{mohit}.
As in CdS, a rotation in the polarization in the reflected beam should also be seen in this case if the incidence direction is not along the surface normal \cite{graham}.

The non-centrosymmetric point groups associated with this order may or may not have
a pseudo-scalar component as well \cite{jc} (Table 2).  Note that for toroidic order, the application of an electric field perpendicular to the toroidal
moment induces a magnetic field perpendicular to both, and vice versa \cite{goulon2}.
A similar `inverse Rashba effect' has been discussed
in Ref.~\onlinecite{fujimoto} for Rashba spin-orbit coupled metals (with $k \times S$ along $z$), where current applied along $y$ leads to a magnetic field along $x$ 
(and vice versa).  For Dresselhaus coupling, one finds a longitudinal response instead \cite{fujimoto}.
Optical activity can be induced as well by currents, the current operator $j$ transforming the same way under parity and time-reversal
as $k$ and $t$.  This has been demonstrated in tellurium, where a current induced (non-reciprocal)
optical rotation is seen in addition to the zero
current (natural) rotation due to the chiral space group \cite{vorobev}.
Ref.~\onlinecite{fu} also mentions the recently claimed ferroelectric metal LiOsO$_3$ \cite{lioso3}.  At the structural
transition, the space group converts from the centrosymmetric R$\bar{3}$c to the non-centrosymmetric R3c.  According to Ref.~\onlinecite{jc},
the latter space group should give rise to vector optical activity, but no pseudo-scalar or pseudo-deviator, and in that sense is similar
to CdS and ZnO, as well as the Weyl semimetal TaAs \cite{TaAs} (Table 2).

Most interesting is $X_3$, which is known in the dichroic literature as a pseudo-deviator \cite{jc}, and again may or may not be observed
depending on the space group.  This operator is largely responsible for x-ray natural circular dichroism (XNCD) \cite{goulon}, and is due to electric dipole (E1) -
electric quadrupole (E2) interference resulting from inversion breaking (circular dichroism being the difference in absorption between left
and right circularly polarized light).  At the K edge, it
again involves orbital as opposed to spin operators.  There are five possible terms (since $Q$ is a traceless symmetric matrix), and depending
on the crystal symmetry, they can form one-dimensional, two-dimensional or three-dimensional representations.  As discussed in Ref.~\onlinecite{fu}, a
double-dimensional representation has been proposed for the tetrahedral pyrochlore Cd$_2$Re$_2$O$_7$ \cite{sergienko} with the associated
Goldstone mode seen by Raman scattering \cite{kendziora}.  Though no optical activity has been reported, second harmonic generation
has been seen \cite{petersen}.

When considering such tensor order in  Cd$_2$Re$_2$O$_7$, note that the double-dimensional representation has as a basis the
two space groups, I4$_1$22 and I$\bar{4}$m2 \cite{sergienko}.  Below the structural ordering temperature, this degeneracy is broken, and the latter
crystal structure has been claimed from x-ray measurements \cite{xray}.  On the other hand, it has been proposed that the former is
realized below a second lower structural transition temperature \cite{second}.  This can be determined by XNCD.  Assuming x-rays
propagating along the $c$ axis of the tetragonally distorted crystal (advantageous as well, since it eliminates any contamination from
linear dichroism), the XNCD signal is zero for the latter crystal structure due to the presence of a mirror plane.  But it can exist for
the former.

\begin{figure}
\includegraphics[width=1.0\hsize]{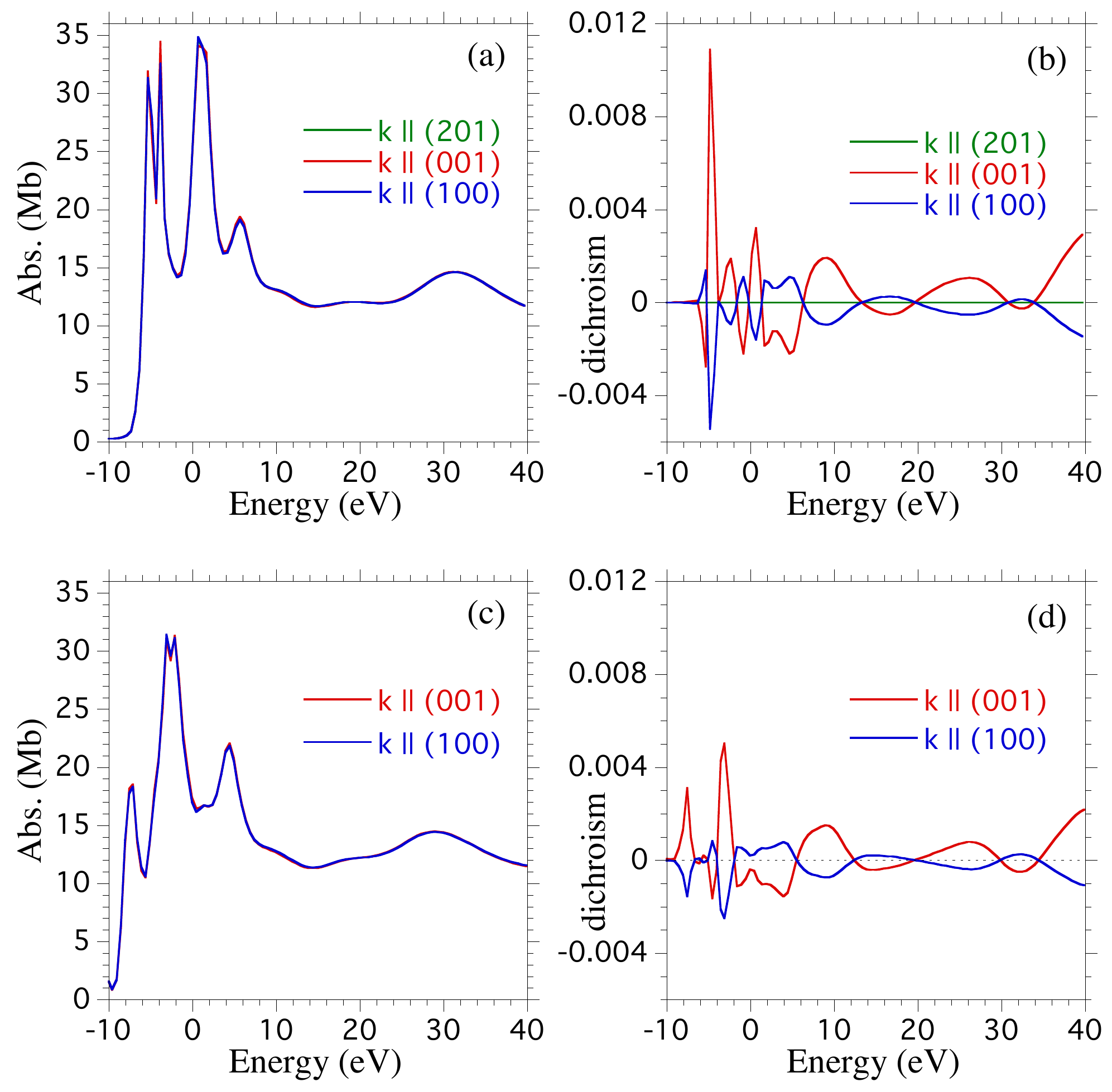}
\caption{(Color online) Calculated X-ray absorption (a) and circular dichroism (b) near the O K edge for Cd$_2$Re$_2$O$_7$ in
the I4$_1$22 space group for $k$
vectors along three directions  - (201), (001) and (100) in tetragonal notation.  Note that (201) corresponds to (111) in cubic notation.
For the dominant O$_1$ sites, the average cluster Fermi energy is at -6.9 eV.
The same is shown in (c) and (d), where now the Re sites are assumed to be in a $d^3p^2$ configuration instead of the $d^5$ configuration
used in (a) and (b).  In this case for the dominant O$_1$ sites, the average cluster Fermi energy is at -8.6 eV.}
\label{fig1}
\end{figure}

This is demonstrated by explicit calculations at the O K edge with the multiple scattering Greens function code FDMNES \cite{joly}
including spin-orbit interactions \cite{so}.
The simulations were done using local density (LDA) atomic potentials (Hedin-Lundqvist 
exchange-correlation function) in a muffin tin approximation that considers multiple scattering of the
photoelectron around the absorbing site \cite{natoli}.  The cluster radius is limited
by the photoelectron lifetime \cite{escape}.  Results shown here are for a cluster radius
of 6 $\AA$, though some calculations were performed for a radius of 7 $\AA$ with similar results.
Atom positions were taken from Huang \etal~\cite{OK2} for the space groups I${\bar4}$m2 (which has no dichroism)
and I4$_1$22.  A Hubbard $U$ of 7 eV \cite{OK2} was used on the Cd sites since the 3d bands sit too high in energy in the LDA.
Note there are two general types of oxygen atoms, six O$_1$ and one O$_2$ per formula unit, $O_1$ having four (three) distinct crystallographic
sites in I${\bar4}$m2 (I4$_1$22), and O$_2$ two (one) in I${\bar4}$m2 (I4$_1$22).

Results are shown in Fig.~1 for various propagation vectors of the x-rays.
The x-ray absorption itself (Fig.~1a) is almost insensitive to the propagation direction since the crystal structure is nearly cubic.
The dichroism, of course, is very sensitive to this (Fig.~1b).  It is essentially zero for a propagation vector of (201) in tetragonal notation,
corresponding to (111) in cubic notation.  This is of note, since all experiments referred to here on Cd$_2$Re$_2$O$_7$ have been conducted on samples
exposing a (111) face.  A sizable dichroism, though, is found for (100) and (001) vectors (right panel).
The absorption at the O K edge for Cd$_2$Re$_2$O$_7$ has been reported by two groups \cite{OK1,OK2}.
Three prominent features are found in the data, somewhat similar to the calculated features in Fig.~1a in the range of -10 eV to 10 eV,
though the lowest energy feature is not split into two as in the calculation, and the middle feature is more peaky in 
the calculation (the experiment having a broad shoulder extending to higher energies).
The correspondence to the experimental data is improved by putting the Re sites into a 5$d^3$ configuration,
promoting the other two 5$d$ electrons into the 6$p$ states (Fig.~1c).
The resulting dichroism signal (Fig.~1d)  is similar to the previous case.
The correspondence could be further improved by increasing the
core hole and photoelectron lifetime broadening \cite{escape}.
Regardless, an appreciable dichroism is predicted, and if looked for, could be used to tell which space group is actually realized
below the upper and lower structural phase transitions.

It may seem puzzling that natural optical activity is being described by products of operators that each break time reversal.
As commented on
by Di Matteo \etal~\cite{sergio}, although XNCD is often claimed to be due to the rank-2 tensor product of the orbital angular momentum and orbital
toroidal moment (row 1 of Table 1) \cite{goulon,marri}, a more natural explanation is due to a quadrupolar arrangement of electric dipoles, known
as an axial toroidal quadrupole (row two of Table 1).
Such is easily demonstrated to be present in V$_2$O$_3$, for instance, due to movements of the oxygen atoms off their
high symmetry locations to form counter propagating
dipole moments on the hexagonal planes above and below the vanadium sites \cite{sergio}.

Finally, more detailed information can often be obtained
from x-ray scattering, since one can obtain information as a function of in-coming and out-going polarization, as well as azimuthal angle.
An additional advantage is that local inversion-breaking effects can also be seen in centrosymmetric space groups due to the differing
phase factor on each atom that arises from Bragg vectors not at the origin \cite{sergio2}.

In summary, natural circular dichroism is a sensitive probe that can be used to identify the inversion-symmetry breaking order parameters suggested
in Ref.~\onlinecite{fu}.  In addition, if time-reversal is broken, other novel types of dichroism can occur as well, including magneto-chiral and non-reciprocal
linear dichroism \cite{goulon}, along with their x-ray scattering analogues \cite{sergio3}.
In general, dichroism and related phenomena should be powerful probes of topological materials,
particularly Weyl semimetals that break either inversion or time-reversal symmetry.
 
The author thanks Liang Fu and Sergio Di Matteo for several helpful discussions.
This work was supported by the Materials Sciences and Engineering
Division, Basic Energy Sciences, Office of Science, US DOE.


\begin{thebibliography}{99}

\bibitem{hosur}
P. Hosur and X.-L. Qi, Phys. Rev. B {\bf 91}, 081106(R) (2015).
\bibitem{zhang}
C. Zhang \etal, arXiv:1504.07698 (2015).
\bibitem{hehl}
F. W. Hehl, Phys. Rev. A {\bf 77}, 022106 (2008).
\bibitem{fu}
Liang Fu, arXiv:1502.00015 (2015).
\bibitem{petersen}
J. C. Petersen \etal, Nature Phys. {\bf 2}, 605 (2006).
\bibitem{fujimoto}
S. Fujimoto, J. Phys. Soc. Japan {\bf 76}, 051008 (2007).
\bibitem{goulon}
J. Goulon \etal, J. Exp. Theor. Phys. {\bf 97}, 402 (2003).
\bibitem{marri}
I. Marri and P. Carra, Phys. Rev. B {\bf 69}, 113101 (2004).
\bibitem{tugushev}
V. M. Dubovik and V. V. Tugushev, Phys. Reports {\bf 187}, 145 (1990).
\bibitem{spaldin}
N. A. Spaldin \etal, Phys. Rev. B {\bf 88}, 094429 (2013).
\bibitem{schmid}
H. Schmid, Ferroelectrics {\bf 252}, 41 (2001).
\bibitem{fiebig}
B. B. Van Aken, J.-P. Rivera, H. Schmid and M. Fiebig, Nature {\bf 449}, 702 (2007).
\bibitem{kubota}
M. Kubota \etal, Phys. Rev. Lett. {\bf 92}, 137401 (2004).
\bibitem{varma}
S. Di Matteo and C. M. Varma, Phys. Rev. B {\bf 67}, 134502 (2003).
\bibitem{gorb}
A. A. Gorbatsevich, O. V. Krivitsky and S. V. Zaykov, Ferroelectrics {\bf 161}, 343 (1994).
\bibitem{zhong}
S. Zhong, J. Orenstein and J. E. Moore, arXiv:1503.02715 (2015).
\bibitem{lovesey2}
S. W. Lovesey and E. Balcar, Phys. Scr. {\bf 81}, 065703 (2010).
\bibitem{sergio}
S. Di Matteo, Y. Joly and C. R. Natoli, Phys. Rev. B {\bf 72}, 144406 (2005).
\bibitem{sergio3}
S. Di Matteo, J. Phys. D: Appl. Phys. {\bf 45}, 163001 (2012).
\bibitem{jc}
J. Jerphagnon and D. S. Chemla, J. Chem. Phys. {\bf 65}, 1522 (1976).
\bibitem{xmcd}
G. Schutz \etal, Phys. Rev. Lett. {\bf 58}, 737 (1987).
\bibitem{bordacs}
S. Bordacs \etal, Nature Phys. {\bf 8}, 734 (2012).
\bibitem{lovesey}
S. W. Lovesey and V. Scagnoli, J. Phys.: Condens. Matter {\bf 21}, 474214 (2009).
\bibitem{marri2}
I. Marri, P. Carra and C. M. Bertoni, J. Phys. A {\bf 39}, 1969 (2006).
\bibitem{barron}
L. D. Barron, {\it Molecular Light Scattering and Optical Activity}, (Cambridge Univ. Pr., Cambridge, 2004).
\bibitem{graham}
E. B. Graham and R. E. Raab, J. Opt. Soc. Am. A {\bf 13}, 1239 (1996).
\bibitem{cds}
E. L. Ivchenko, S. A. Permogorov and A. V. Selkin, Solid State Comm. {\bf 28}, 345 (1978).
\bibitem{zno}
J. Goulon \etal, J. Phys: Condens. Matter {\bf 19}, 156201 (2007).
\bibitem{mohit}
M. Kargarian, M. Randeria and N. Trivedi, arXiv:1503.00012 (2015).
\bibitem{goulon2}
J. Goulon \etal, Phys. Rev. Lett. {\bf 88}, 237401 (2002).
\bibitem{vorobev}
L. E. Vorobev \etal, JETP Lett. {\bf 29}, 441 (1979).
\bibitem{lioso3}
Y. Shi \etal, Nature Matls. {\bf 12}, 1024 (2013).
\bibitem{TaAs}
H. Weng \etal, Phys. Rev. X {\bf 5}, 011029 (2015).
\bibitem{sergienko}
I. A. Sergienko and S. H. Curnoe, J. Phys. Soc. Japan {\bf 72}, 1607 (2003).
\bibitem{kendziora}
C. A. Kendziora \etal, Phys. Rev. Lett. {\bf 95}, 125503 (2005).
\bibitem{xray}
M. T. Weller \etal, Dalton Trans. 3032 (2004).
\bibitem{second}
J. I. Yamaura and Z. Hiroi, J. Phys. Soc. Japan {\bf 71}, 2598 (2002).
\bibitem{joly}
Y. Joly, Phys. Rev. B {\bf 63}, 125120 (2001). The FDMNES program can be 
downloaded at http://neel.cnrs.fr/spip.php?rubrique1007
\bibitem{so}
Y. Joly, O. Bunau, J. E. Lorenzo, R. M. Galera, S. Grenier and B. Thompson, J. Phys.: Conf. Ser. {\bf 190}, 012007 (2009).
\bibitem{natoli}
C. R. Natoli, Ch. Brouder, Ph. Sainctavit, J. Goulon, Ch. Goulon-Ginet and A. Rogalev, Eur. Phys. J. B {\bf 4}, 1 (1998).
\bibitem{escape}
The results presented involve a convolution of the calculated spectrum with 
both a core hole (0.2 eV) and a photoelectron
inverse lifetime, with the latter having a strong energy dependence (at high energies, 15 eV, with a
midpoint value at 30 eV above the Fermi energy) \cite{so}.  Core hole widths tabulated
in the FDMNES program are based on the x-ray literature \cite{so}.
\bibitem{OK2}
S.-W. Huang \etal, J. Phys.: Condens. Matter {\bf 21}, 195602 (2009).
\bibitem{OK1}
A. Irizawa \etal, J. Phys. Soc. Japan {\bf 75}, 094701 (2006).
\bibitem{sergio2}
S. Di Matteo \etal, Phys. Rev. Lett. {\bf 91}, 257402 (2003).

\end{thebibliography}
\end{document}